\documentclass[12pt]{article}

\usepackage{amssymb}
\usepackage{amsmath,amsthm, mathtools,commath}
\usepackage{graphics, color}

\usepackage{graphicx}
\usepackage{epsfig}
\usepackage{makeidx}
\usepackage{kotex}
\usepackage{fullpage}
\usepackage{bm}
\usepackage{float}
\usepackage{url}
\usepackage{enumitem}
\usepackage{enumitem}

\usepackage[round]{natbib}

\usepackage{enumitem}

\newtheorem{theorem}{Theorem}
\newtheorem{lemma}[theorem]{Lemma}
\newtheorem{corollary}[theorem]{Corollary}
\newtheorem{proposition}[theorem]{Proposition}
\newtheorem{remark}{Remark}

\usepackage[round]{natbib}

\usepackage{amsmath, amssymb, amsfonts, amsthm, bm, commath, bbm}
\usepackage{booktabs, arydshln}
\usepackage{comment}

\usepackage{tikz}
\usetikzlibrary{patterns}
\usetikzlibrary{calc}
\usepackage{standalone}
\usetikzlibrary{arrows.meta,positioning}
\usetikzlibrary{arrows.meta,positioning,shadows.blur,fit,backgrounds}

\newcommand{\Ep}{\mathbb{E}_p}
\newcommand{\Varp}{\mathbb{V}\mathrm{ar}_p}

\newcommand{\Pp}{\mathbb{P}_p}

\begin{document}

\title{{\it S}ample-split {\it REG}ression SREG: A Robust Estimator for High-Dimensional Survey Data}


\author{Yonghyun Kwon \and Shu Yang \and Jae Kwang Kim }

\maketitle

\begin{abstract} 
Model-assisted regression estimation is fundamental in survey sampling for incorporating auxiliary information. However, when the auxiliary dimension $p$ grows with the sample size $n$, the standard {\it G}eneralized {\it REG}ression (GREG) estimator can exhibit non-negligible bias under informative sampling, even when the working model is correctly specified. This failure stems from the ``double use'' of sampled outcomes—simultaneously for fitting the regression and for forming the residual correction. We propose a {\it S}ample-split {\it REG}ression (SREG) estimator based on $K$-fold cross-fitting that eliminates this bias by pairing each unit's residual with an out-of-fold prediction. The resulting estimator is first-order equivalent to the oracle difference estimator under a weak prediction-norm consistency requirement, without requiring $\sqrt{n}$-consistent estimation of regression coefficients. We establish asymptotic normality and prove consistency of a variance estimator based on cross-fitted residuals. The key conditional fluctuation assumption is verified for simple random, stratified, and rejective sampling. Simulations demonstrate that SREG effectively removes high-dimensional bias while maintaining competitive efficiency.
\end{abstract}


\baselineskip .3in

\newpage

\section{Introduction}

Model-assisted estimation has long been a central tool in finite population
survey sampling when reliable auxiliary information is available.
In many large-scale surveys, the study variable $y$ is observed only
for sampled units, while a vector of auxiliary variables $x$ is known
for all population units (e.g., from administrative records or frame
data). Regression-type estimators exploit this structure by combining
(i) model-based predictions for the full population and (ii) a design-weighted
correction that restores design consistency. A prominent example is
the generalized regression (GREG), or model-assisted regression estimator
\citep{deville1992calibration,fuller2002,breidt2017model}, which \textcolor{black}{is
robust to model misspecification} and can deliver substantial gains
in efficiency over the Horvitz--Thompson (HT) estimator when the
working outcome regression captures meaningful signal.

The classical asymptotic theory for model-assisted regression estimators
typically assumes a fixed and moderate-dimensional auxiliary vector.
Contemporary  applications, however, increasingly involve high-dimensional
auxiliary information: rich paradata, administrative histories, transactional
summaries, or constructed features derived from linked data sources.
In such settings, the dimension $p$ of $x$ can grow with the sample
size $n$, and it is common for $p$ to be of the same order as $n$,
or even larger.  In this regime, naively augmenting the regression with many auxiliary variables can degrade the estimator's design-based behavior \citep{goga2025}. Moreover, under informative sampling  \citep{pfe99}, the bias induced by fitting a high-dimensional regression within the sample can be non-negligible and may not vanish at the usual $\sqrt{n}$ rate.

A key mechanism behind this failure is the “double use” of the sampled
outcomes: the same 
$y$ values are used both to fit the regression and to form the HT residual correction. When 
$p$ is fixed, the resulting feedback is asymptotically negligible. When $p$ grows with $n$, however, the self-influence of each sampled outcome on its own fitted value accumulates; under informative sampling this can translate into first-order bias. In particular, when 
$p/n$  does not vanish, the ordinary GREG estimator can fail to be asymptotically equivalent to the oracle difference estimator, even under correct specification \citep{ta2020generalized, chauvet2022, dagdoug2023modelassisted}.

This paper proposes a simple remedy based on sample splitting and cross-fitting \citep{chernozhukov2018double, bates2024cross}. We randomly partition the finite population into $K$ folds, independently of the sampling design. For each fold $k$, we fit the working regression using only sampled observations outside fold $k$, then form a fold-specific regression estimator using out-of-fold predictions. The resulting $K$-fold sample-split regression estimator (SREG) uses all sampled observations while maintaining out-of-fold honesty for every residual term.

Bias correction via resampling has precedent in design-based survey sampling; \citet{hartley1954unbiased}, \citet{mickey1959some}, and \citet{williams1961} used sample splitting for unbiased ratio and regression estimation under simple random sampling.  More recently, \citet{sande2021design} proposed a subsampling Rao--Blackwell method for unbiased regression estimation, and \citet{stefan2024jackknife} applied the jackknife procedure to correct the bias of the GREG estimator. \citet{eustache2025high} developed resampling methods for bias-corrected variance estimation in high-dimensional settings. However, these approaches largely focus on linear working models and do not address the broader question of accommodating general prediction rules under high-dimensional asymptotics. 

The key advantage of sample splitting is that controlling the remainder no longer requires $\sqrt{n}$-consistent estimation of regression coefficients. Instead, it suffices that the out-of-fold predictor be consistent in an $L_2$  (prediction-norm) sense over the finite population. 
\textcolor{black}{This requirement is substantially
weaker and accommodates }ordinary least squares, weighted least squares,
penalized regression methods such as lasso, or more general machine-learning
predictors, provided the out-of-fold prediction error converges appropriately.

We develop a design-based asymptotic theory for the proposed estimator
under a standard finite-population sequence framework. The main contributions of this paper are: (i) a simple sample-split regression estimator that eliminates high-dimensional bias under informative sampling; (ii) design-based asymptotic theory requiring only prediction-norm consistency rather than $\sqrt{n}$-consistent coefficient estimation; (iii) verification of the key conditional fluctuation condition for SRSWOR, stratified SRSWOR, and rejective sampling; and (iv) a consistent variance estimator based on cross-fitted residuals. 
\cite{lu2025debiased} develops the design-based theory for causal inference with independent treatment assignment; our contributions develop this theory for complex survey designs, addressing an important gap.

A simulation study illustrates the practical implications of the theory.
In settings with many auxiliary variables and an informative stratified
design, the ordinary GREG estimator can exhibit substantial bias,
while the proposed sample-split versions remove this bias at the cost
of a modest increase in variance.  The proposed variance estimator
exhibits negligible relative bias and delivers coverage closer to
nominal levels than variance estimators paired with the biased ordinary
regression estimator under informative sampling.

The remainder of the paper is organized as follows. Section \ref{sec:basic}
introduces the basic setup and clarifies why the standard model-assisted
regression estimator can fail in high dimension under informative
designs. Section \ref{sec:method} presents the proposed $K$-fold
sample-split (cross-fitted) regression estimator. Section \ref{sec:theory}
develops the main design-based asymptotic theory and variance estimation
results. Section \ref{sec:Controlling-the-remainder} verifies the
key conditional fluctuation condition for several widely used sampling
designs. Section \ref{sec:Simulation-Study} reports simulation results. Concluding Remarks are made in Section 7. 
Proofs and a real data application are presented in the supplementary material.

\section{Basic Setup}\label{sec:basic}

\subsection{Finite-population framework and target estimand}

Let $U_{N}=\{1,\ldots,N\}$ denote a finite population of size $N$.
For each unit $i\in U_{N}$, let $y_{i}$ be the study variable and
let $\bm x_{i}\in\mathbb{R}^{p}$ be a $p$-dimensional vector of auxiliary
variables. The target parameter is the population total $T=\sum_{i\in U_{N}}y_{i}$. 
A probability sample $A\subset U_{N}$ of size $n=|A|$ is selected
according to a sampling design $\Pp$. Let $I_{i}=\mathbbm{1}(i\in A)$
be the sample membership indicator, and let $\pi_{i}=\Pp(I_{i}=1)$ and $\pi_{ij}=\Pp(I_{i}=1,I_{j}=1)$ 
denote the first- and second-order inclusion probabilities, respectively.
We assume that the auxiliary variables $\{ {\bm x}_{i} \}_{i \in U_N} $ are
observed for all units in the population, while $\{y_{i}\}_{i\in U_{N}}$
are observed only for sampled units $i\in A$.

Model-assisted regression estimation is motivated by a working superpopulation model of the form
\begin{equation}
y_{i}=m(\bm x_{i};\beta_{0})+e_{i},\label{eq:working_model}
\end{equation}
where $m(\cdot;\beta)$ is a specified functional form (e.g., linear) and $\beta_{0}$ is a population-level parameter. When the model is correctly specified, $m(\bm x_{i};\beta_{0})=\mathbb{E}(y_{i}\mid \bm x_{i})$ and $\mathbb{E}(e_{i}\mid \bm x_{i})=0$. We assume that $m(x;\beta)$ is continuously
differentiable in $\beta$ in a neighborhood of $\beta_{0}$, and
we write $\dot{m}(x;\beta)=\partial m(x;\beta)/\partial\beta$.

If the mean function $m(\bm x_{i};\beta_{0})$ were known for all $i\in U_{N}$,
a natural benchmark is \textbf{the oracle difference estimator} 
\begin{equation}
\widehat{T}_{\mathrm{diff}}=\sum_{i\in U_{N}}m(\bm x_{i};\beta_{0})+\sum_{i\in A}\frac{1}{\pi_{i}}\Big\{ y_{i}-m(\bm x_{i};\beta_{0})\Big\}.\label{eq:oracle_diff}
\end{equation}
This estimator has the usual HT form applied to the oracle residuals
$e_{i}$,  
plus the (known) population total
of the oracle predictions.

\subsection{Ordinary model-assisted regression (GREG)}

In practice, $\beta_{0}$ is unknown and must be estimated from the
sample. Let $\widehat{\beta}$ be an estimator of $\beta_{0}$ constructed
using the sampled observations (e.g., OLS/WLS in a working linear
model). The standard \textbf{GREG} estimator of $T$ is 
\begin{equation}
\widehat{T}_{\mathrm{reg}}=\sum_{i\in U_{N}}m(\bm x_{i};\widehat{\beta})+\sum_{i\in A}\frac{1}{\pi_{i}}\Big\{ y_{i}-m(\bm x_{i};\widehat{\beta})\Big\}.\label{eq:greg}
\end{equation}
When the auxiliary dimension $p$ is fixed and $\widehat{\beta}$
is $\sqrt{n}$-consistent, standard arguments imply that $\widehat{T}_{\mathrm{reg}}$
is first-order equivalent to the oracle difference estimator \eqref{eq:oracle_diff},
and therefore inherits the same first-order design-based behavior.

Modern survey applications, however, often involve \textbf{high-dimensional
auxiliary information}, and it is natural to allow the dimension $p=p_{n}$
to grow with the sample size $n$. In this regime, the first-order
behavior of the regression estimator can change substantially, particularly
under informative sampling.

\subsection{Diagnosing the high-dimensional failure}

To understand why the standard model-assisted regression estimator can fail in high dimensions, it is helpful to first identify the structural source of the problem. In the standard GREG construction, the same sampled outcomes $y_i$ are used twice:
\begin{enumerate}
    \item to estimate the regression function through $\hat{\beta}$, and
    \item to form the HT residual correction.
\end{enumerate}
When the auxiliary dimension is fixed, the resulting self-influence is asymptotically negligible. In high dimensions, however, each sampled unit can exert non-trivial influence on its own fitted value. Under informative sampling, this self-influence interacts with unequal inclusion probabilities and accumulates into a first-order distortion.

Importantly, this failure is not merely a variance inflation or overfitting issue. Even when the working regression model is correctly specified, the ordinary model-assisted regression estimator can lose its oracle-equivalence property solely due to the reuse of outcomes in a high-dimensional setting.

To make this precise, compare \eqref{eq:greg} to the oracle benchmark \eqref{eq:oracle_diff}.
A direct subtraction yields 
\[
N^{-1}\big(\widehat{T}_{\mathrm{reg}}-\widehat{T}_{\mathrm{diff}}\big)=-\frac{1}{N}\sum_{i\in U_{N}}\Big(\frac{I_{i}}{\pi_{i}}-1\Big)\Big\{ m(\bm x_{i};\widehat{\beta})-m(\bm x_{i};\beta_{0})\Big\}.
\]
By a mean-value expansion, 
\[
m(\bm x_{i};\widehat{\beta})-m(\bm x_{i};\beta_{0})=\dot{m}(\bm x_{i};\beta^{\ast})^{\top}(\widehat{\beta}-\beta_{0}),
\]
where $\beta^{\ast}$ lies on the line segment between $\widehat{\beta}$
and $\beta_{0}$, so that 
\[
N^{-1}\big(\widehat{T}_{\mathrm{reg}}-\widehat{T}_{\mathrm{diff}}\big)=\frac{1}{N}\left\{ \sum_{i\in U_{N}}\Big(\frac{I_{i}}{\pi_{i}}-1\Big)\dot{m}(\bm x_{i};\beta^{\ast})^{\top}\right\} (\widehat{\beta}-\beta_{0}).
\]
Consequently, the discrepancy between the regression estimator and
the oracle estimator depends directly on the interaction between (i)
estimation error in $\widehat{\beta}$ and (ii) the design-weighted
fluctuation term $\sum_{i\in U_{N}}(I_{i}/\pi_{i}-1)\dot{m}(\bm x_{i};\beta^{\ast})^{\top}$.

Under high-dimensional asymptotics, this interaction can accumulate. 
In particular, under a linear working model, \citet{ta2020generalized} showed that the asymptotic equivalence may fail when $p/n \not \to 0$.
Under more restrictive regularity conditions, \citet{chauvet2022} show that
\begin{equation}
\sqrt{n}\,N^{-1}\big(\widehat{T}_{\mathrm{reg}}-\widehat{T}_{\mathrm{diff}}\big)=O_{p}\!\left(\frac{p}{\sqrt{n}}\right)+O_{p}\left(\frac{p^{2}}{n}\right),
\end{equation}
and hence
\begin{equation}
\sqrt{n}\,N^{-1}\big(\widehat{T}_{\mathrm{reg}}-T\big)=\sqrt{n}\,N^{-1}\big(\widehat{T}_{\mathrm{diff}}-T\big)+O_{p}\!\left(\frac{p}{\sqrt{n}}\right)+O_{p}\left(\frac{p^{2}}{n}\right).\label{eq:non_equiv}
\end{equation}
Therefore, if $p^{2}/n\to\lambda>0$, the extra term in \eqref{eq:non_equiv}
does not vanish in general, and $\widehat{T}_{\mathrm{reg}}$ need
not be asymptotically equivalent to the oracle difference estimator.
In particular, when the sampling design is informative, the additional
term can manifest as a non-negligible bias in the limit distribution
of $\widehat{T}_{\mathrm{reg}}$.

\begin{remark}[Design-based bias under informative sampling]
\label{rem:greg_bias}
Even under correct specification, the ordinary GREG estimator is not design-unbiased
under informative sampling. For the linear case, the design-based bias takes the form
\[
\frac{1}{N}\mathbb E_p\!\left(\widehat{T}_{\mathrm{reg}}-T\right)
=
-\frac{1}{N^2}\sum_{i\in U_N}\sum_{j\in U_N}
\frac{\Delta_{ij}}{\pi_i\pi_j}\,
e_{i,N}\,
\bm x_j^\top Q_N^{-1}\bm x_i
+o(n^{-1}),
\]
where $\Delta_{ij}=\pi_{ij}-\pi_i\pi_j$, and $Q_N$ and $e_{i, N}$ are defined in the supplementary material, as derived in Chapter 7 of \citet{cochran1977sampling} under simple random sampling.
For fixed $p$, this bias is $O(n^{-1})$ and hence asymptotically negligible relative to sampling variability.
However, it can grow with $p$,
so the bias can accumulate and become non-negligible when $p$ is large.
\end{remark}

The preceding analysis points to a structural remedy: decouple regression fitting from residual correction so that, for each sampled unit, the prediction used in its residual term is constructed without using that unit's own outcome. Section~\ref{sec:method} implements
this idea via a random $K$-fold partition of the population and a
cross-fitted (sample-split) regression estimator that preserves the
difference-estimator form while eliminating the leading high-dimensional
self-influence that drives \eqref{eq:non_equiv}.

\section{Proposal: Sample-Split Regression Estimation}

\label{sec:method}

Section \ref{sec:basic} shows that the ordinary model-assisted regression
estimator can fail to be first-order equivalent to the oracle difference
estimator when the auxiliary dimension grows with the sample size
under an informative sampling design. The core difficulty is the ``double
use'' of the same sampled outcomes: the fitted regression function
is trained on the sample, and the same sample is then reused to form
the HT residual correction. When the fitted regression is high-dimensional
and the design is informative, the resulting extra term need not be
asymptotically negligible.

To address this issue, we propose a \emph{sample-split / cross-fitted
regression estimator} that pairs each sampled unit's outcome with
an \emph{out-of-fold} prediction constructed without using that unit's
outcome in training. The construction is closely related to cross-fitting
in semiparametric inference, but it is implemented here in a design-based
finite-population framework.

\subsection{Sample splitting and cross-fitting}

We first split the finite population into $K$ disjoint folds. Independently of the sampling design and of the finite-population values $\{(y_{i},\bm x_{i})\}_{i\in U_{N}}$, assign each unit $i\in U_{N}$ a fold label $\kappa_{i}\in\{1,\dots,K\}$ drawn independently from the discrete uniform distribution on $\{1,\ldots,K\}$. Define the population folds and
their sample intersections by 
\[
U_{k}:=\{i\in U_{N}:\kappa_{i}=k\},\qquad A_{k}:=A\cap U_{k},\qquad N_{k}:=|U_{k}|,\qquad n_{k}:=|A_{k}|.
\]
Let $k(i)$ denote the unique fold index such that $i\in U_{k(i)}$.
The population total can be expressed as $T=\sum_{k=1}^{K}T_{k}$ where $T_{k}=\sum_{i\in U_{k}}y_{i}$.

For each fold $k$, we estimate $\beta_{0}$ in the model \eqref{eq:working_model}
using only the sample
observations \emph{excluding} fold $k$, i.e.\ using $\{(y_{i},\bm x_{i}):i\in A\setminus A_{k}\}$,
and denote the resulting estimator by $\hat{\beta}^{(-k)}$. This
yields the out-of-fold predictor 
$\hat{m}^{(-k)}(x):=m\bigl(x;\hat{\beta}^{(-k)}\bigr),\qquad\hat{m}_{i}^{(-k)}:=\hat{m}^{(-k)}(\bm x_{i})$.
The key ``honesty'' property is that for $i\in A_{k}$, the prediction
$\hat{m}^{(-k)}(\bm x_{i})$ is constructed without using $y_{i}$.

\subsection{Fold-Wise regression estimators and aggregation}

For each fold $k$, define the fold-wise sample-split regression
estimator 
\begin{equation}
\hat{T}_{k,\mathrm{reg}}^{(-k)}:=\sum_{i\in U_{k}}\hat{m}_{i}^{(-k)}\;+\;\sum_{i\in A_{k}}\frac{1}{\pi_{i}}\Bigl\{ y_{i}-\hat{m}_{i}^{(-k)}\Bigr\}.\label{eq:fold-est}
\end{equation}
The proposed $K$-fold sample-split regression estimator of the overall
population total $T:=\sum_{i\in U_{N}}y_{i}$ is 
\begin{equation}
\hat{T}_{{\mathrm{SREG}}}:=\sum_{k=1}^{K}\hat{T}_{k,\mathrm{reg}}^{(-k)}.\label{eq:TSS}
\end{equation}

\begin{figure}[t]
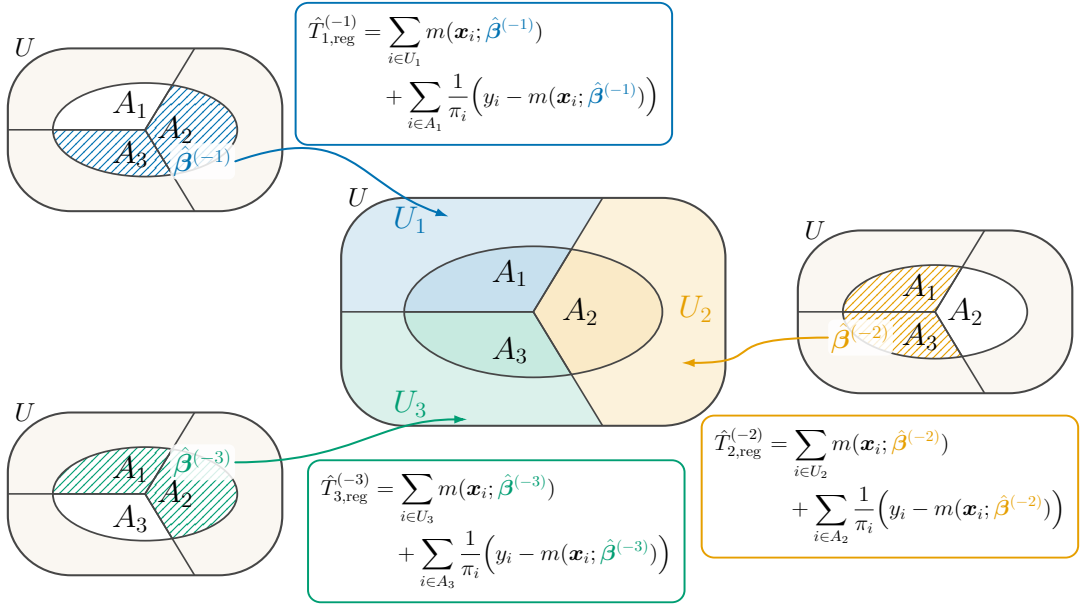

\centering
  \includestandalone[width=15cm,height=8cm]{fig/SREGdiagram}
\caption{Illustration
of the SREG estimator for $K=3$.}
\label{fig:Illustration-of-the}
\end{figure}

Figure~\ref{fig:Illustration-of-the} illustrates the construction for $K=3$ folds: for each fold, the regression is fitted using sampled units outside that fold, and the foldwise total $T_k$ is estimated using sampled units inside that fold. The $K$-fold estimator \eqref{eq:TSS} aggregates these fold-wise estimates, using all sampled units for estimation while maintaining out-of-fold honesty. The algorithm is summarized in Figure~\ref{fig:algorithm}.

\subsection{Exact decomposition relative to the oracle estimator}
To understand the proposed sample-split estimator, recall the oracle
difference estimator 
\begin{equation*}
\widehat{T}_{\mathrm{diff}}:=\sum_{i\in U_{N}}m_{i}\;+\;\sum_{i\in A}\frac{1}{\pi_{i}}(y_{i}-m_{i}),
\end{equation*}
where $m_{i}=m(\bm x_{i};\beta_{0})$. The SREG estimator \eqref{eq:TSS}
can be viewed as a regression/difference estimator that replaces the
unknown $m_{i}$ with out-of-fold predictions $\widehat{m}_{i}^{(-k(i))}$.
Note that the discrepancy between $\hat{T}_{\mathrm{SREG}}$ and the
oracle difference estimator $\hat T_{\rm diff}$ admits the exact decomposition 
\begin{equation}
\widehat{T}_{\mathrm{SREG}}-\widehat{T}_{\mathrm{diff}}=\sum_{k=1}^{K}\sum_{i\in U_{k}}\Bigl(1-\frac{I_{i}}{\pi_{i}}\Bigr)\Bigl\{\widehat{m}_{i}^{(-k)}-m_{i}\Bigr\}.\label{eq:remainder-sec3}
\end{equation}
The right-hand side is an HT fluctuation applied to the \emph{out-of-fold}
prediction error $\widehat{m}_{i}^{(-k)}-m_{i}$.  Define 
\begin{eqnarray*}
r_{N,k}\equiv \frac{1}{N_{k}}\left(\widehat{T}_{k,{\rm reg}}^{(-k)}-\widehat{T}_{k,{\rm diff}}\right) & = & \frac{1}{N_{k}}\sum_{i\in U_{k}}\left(1-\frac{I_{i}}{\pi_{i}}\right)\left\{ \widehat{m}^{(-k)}(\bm x_{i})-m(\bm x_{i})\right\} ,
\end{eqnarray*}
where $\widehat{m}^{(-k)}(\cdot)=m(\cdot;\hat{\bm \beta}^{(-k)})$. To build intuition, consider Poisson sampling, where $\{I_i\}_{i\in U_N}$ are mutually independent. 
Because $\widehat{m}^{(-k)}$ is constructed without using $\{I_i, : i \in U_k\}$, the prediction $\widehat{m}_{i}^{(-k)}$ is independent of $I_i$, and
\begin{equation*}
\Ep[\{1-I_{i}/\pi_{i}\}\{\widehat{m}^{(-k)}(\bm x_{i})-m(\bm x_{i})\}\mid\widehat{m}^{(-k)}(\bm x_{i})]=0 
\end{equation*}
if $i \in U_k$. Thus, we have 
\begin{eqnarray}
\Ep(r_{N,k}^{2}) & \leq & \frac{C}{n}\max_{i \in U_k}\Ep\{\hat{m}^{(-k)}(\bm x_i)-m(\bm x_i)\}^{2}\label{bdd}
\end{eqnarray}
for some constant $C$ if $\pi_{i}^{-1}$ is uniformly bounded. Therefore, $\Ep(r_{N,k}^{2})=o(n^{-1})$
if $\max_{i \in U_k}\Ep\{\widehat{m}^{(-k)}(\bm x_i)-m(\bm x_i)\}^{2}=o(1)$, namely, the
prediction norm of $\widehat{m}^{(-k)}(\cdot)$ uniformly converges.

Crucially, because $\widehat{m}^{(-k)}$ is constructed without using
$\{y_{i}:i\in A_{k}\}$, the leading ``self-influence'' term that
drives the high-dimensional bias in the regression estimator
is eliminated. As a result, the remainder in \eqref{eq:remainder-sec3}
can be controlled under a \emph{prediction-norm} requirement for $\widehat{m}^{(-k)}$,
without requiring $\sqrt{n}$-consistency of $\hat{\beta}^{(-k)}$
in high dimension. 

Section \ref{sec:theory} formalizes the above argument and provides
a design-based asymptotic theory for $\hat{T}_{\mathrm{SREG}}$ and
its variance estimator, including first-order equivalence to the oracle
difference estimator, asymptotic normality, and variance-estimator
consistency.

\section{Design-Based Asymptotic Theory}

\label{sec:theory}

To formally discuss asymptotic theory of the proposed method, consider
a sequence of finite populations $\{U_{N}\}_{N\ge1}$ as in \citet{isaki1982survey}.
We state conditions in a design-based asymptotic regime where $N\to\infty$,
$n\to\infty$, and $n/N\to f\in(0,1)$. All design-based expectations
$\Ep(\cdot)$ are conditional on the finite population $\{(\bm x_{i},y_{i});i\in U_{N}\}$,
thus $(\bm x_{i},y_{i})$ are treated as fixed under $\Pp$.

\begin{enumerate}[label=(A\arabic*), leftmargin=*]
\item \label{A:pos} \textbf{Bounded sampling weights.} There exist constants $0<\pi_{\min}\le\pi_{\max}<1$ such that  $\pi_{\min}\le\pi_{i}\le\pi_{\max},\qquad \forall i \in U_N,\ \forall N$. 
\item \label{A:depend} \textbf{Controlled second-order inclusion dependence.} Let $\Delta_{ij}:=\pi_{ij}-\pi_{i}\pi_{j}$ for $i,j\in U_{N}$. Assume
there exists a constant $C_{\Delta}<\infty$ such that
\begin{equation}
\max_{i\in U_{N}}\ \sum_{j\in U_{N}}|\Delta_{ij}|
\ \le\ C_{\Delta}\,\frac{n}{N}.
\label{eq:A2Delta_rowsum}
\end{equation}


\item \label{A:moments} \textbf{Finite second moments.}
\[
\frac{1}{N}\sum_{i\in U_{N}}y_{i}^{2}=O(1),
\qquad
\frac{1}{N}\sum_{i\in U_{N}}m_{i}^{2}=O(1).
\]

\item \label{A:CF} \textbf{Cross-fitting (honesty).} For each fold $k$, the fitted predictor $\widehat{m}^{(-k)}(\cdot)$
is measurable with respect to the out-of-fold units
$\{I_{i}:i\in U_{N}\setminus U_{k}\}$
and does not use $\{y_{i}: i\in A_{k}\}$ in training. For each $i \in U_N$, a fold label $\kappa_{i}\in\{1,\dots,K\}$ is drawn independently from the discrete uniform distribution on $\{1,\ldots,K\}$, and the number of folds satisfies $\limsup_{N\to \infty} K < \infty$.

\item \label{A:normcons} \textbf{Prediction-norm consistency on the finite population.} Define the fold-wise prediction error
\[
\|\widehat{m}^{(-k)}-m\|_{N,k}^{2}
:=\frac{1}{N_{k}}\sum_{i\in U_{k}}
\{\widehat{m}_{i}^{(-k)}-m_{i}\}^{2}.
\]
Assume
\[
\max_{1\le k\le K}\ 
\|\widehat{m}^{(-k)}-m\|_{N,k}
=o_{p}(1).
\]

\end{enumerate}
Assumption \ref{A:pos} is common in the survey sampling literature.  
To derive asymptotic properties, Assumption \ref{A:depend} controls sampling dependence in the spirit of \citet{breidt2008endogenous} and is weaker than Assumption A6 of \citet{breidt2000local} and Assumption A5 of \citet{robinson1983asymptotic}. Assumption \ref{A:moments} is also common and imposes a mild restriction on second-order finite-population moments.  
We further impose Assumption \ref{A:CF} to prevent double-dipping in the sense of \citet{chernozhukov2018double}. Assumption \ref{A:normcons} is a \emph{prediction} requirement: it does not require $\sqrt{n}$-consistency of the nuisance estimator $\widehat{\beta}^{(-k)}$. Similar prediction-type conditions appear in \citet{toth2011building} for regression trees and in \citet{dagdoug2023model} for random forests without sample splitting. This prediction error condition on nuisance parameter is standard in the double/debiased machine learning framework and is sufficient for valid inference when combined with Neyman orthogonality.

Define the remainder $R_{N}:=\widehat{T}_{\mathrm{SREG}}-\widehat{T}_{\mathrm{diff}}$. 
A direct algebraic simplification yields 
\begin{equation}
R_{N}=\sum_{k=1}^{K}\sum_{i\in U_{k}}\Big(1-\frac{I_{i}}{\pi_{i}}\Big)\{\widehat{m}_{i}^{(-k)}-m_{i}\}.\label{eq:remainder_decomp}
\end{equation}
Thus, the discrepancy between $\widehat{T}_{\mathrm{SREG}}$ and $\widehat{T}_{\mathrm{diff}}$
is a HT fluctuation applied to the \emph{out-of-fold} prediction error.
For each fold $k=1,\dots,K$, define the training sigma-field 
$$\mathcal{G}_{k}:=\sigma\Big(\set{\kappa_i}_{i \in U_N}, \{I_{j}:j\in U_{N}\setminus U_{k}\}\Big).$$ 
For each fold $k$, define the fold-wise remainder 
\[
R_{N,k}:=\sum_{i\in U_{k}}\Big(1-\frac{I_{i}}{\pi_{i}}\Big)\Big(\widehat{m}_{i}^{(-k)}-m_{i}\Big),\qquad\text{so that }R_{N}=\sum_{k=1}^{K}R_{N,k}.
\]
Under Poisson sampling, we can establish the following lemma.

\begin{lemma}[Mean-square control of the Sample-split remainder under Poisson sampling]\label{lem:4.1}
Assume \ref{A:pos} and \ref{A:CF}. Under Poisson sampling, for each $k=1,\ldots,K$,
\begin{equation}
\Ep\left(R_{N,k}\Big|\ \mathcal{G}_{k}\right)=0.\label{eq:res1}
\end{equation}
Consequently, 
\begin{equation}
\Ep\!\left[\left(\frac{R_{N}}{N}\right)^{2}\right]\ \le\ \frac{C}{n}\cdot \Ep\sbr{ \max_{1\le k\le K}\ \|\widehat{m}^{(-k)}-m\|_{N,k}^{2}},\label{eq:overall}
\end{equation}
for a constant $C<\infty$ and for all large enough $N$. \end{lemma}

Lemma \ref{lem:4.1} presents the upper bound of the remainder term.
Combined with \ref{A:normcons}, the upper bound of $R_N / N$ is $o_{p}(n^{-1/2})$, which is
negligible. However, Lemma \ref{lem:4.1} is justified under Poisson
sampling only. Note $\widehat{m}^{(-k)}$ is trained using the sample
observations in $A/A_{k}$. Therefore, $\hat{m}^{(-k)}$ is a function
of inclusion indicators $\{I_{j};j\notin U_{k}\}$ and sampled $y_{j}$'s.
If the sample design enforces a fixed sample size or otherwise couples
inclusion indicators, then $\Ep\left(I_{i}\mid\hat{m}^{(-k)}\right)\neq\pi_{i}$
for $i\in U_{k}$. So (\ref{eq:res1}) is not generally correct unless
we assume a design with appropriate conditional independence properties.

To achieve (\ref{eq:overall}) without requiring (\ref{eq:res1}),
we introduce the following assumption.

\begin{enumerate}[label=(A\arabic*$^*$), leftmargin=*, start = 2] 
\item \label{A:fluct} \textbf{Fold-wise conditional HT fluctuation bound.}
For any $0<\eta<1$, there exists a constant $C_{\eta}$ and $N_\eta$ such
that with probability at least $1-\eta$ for all $N \geq N_\eta$, for each
$k=1,\dots,K$ and any $\mathcal{G}_{k}$-measurable array $\{a_{i}^{(k)}:\ i\in U_{k}\}$,
\begin{equation}
\Ep\Bigg[\Bigg\{\frac{1}{N_{k}}\sum_{i\in U_{k}}\Big(\frac{I_{i}}{\pi_{i}}-1\Big)a_{i}^{(k)}\Bigg\}^{2}\ \Bigg|\ \mathcal{G}_{k}\Bigg]\ \le\ \frac{C_{\eta}}{n}\cdot\frac{1}{N_{k}}\sum_{i\in U_{k}}\big(a_{i}^{(k)}\big)^{2}.\label{A2*}
\end{equation}
\end{enumerate}

Roughly speaking, (\ref{A2*}) means that the conditional expectation
of $R_{N,k}^{2}$ is dominated by the conditional variance of $R_{N,k}$. \textcolor{black}{Assumption \ref{A:fluct} may appear abstract, as it is
generally infeasible to characterize it for arbitrary sampling designs;
doing so would require deriving design properties conditional on realizations
of partial samples. Nevertheless, the assumption is not unduly restrictive. When $K = 1$, \ref{A:depend} implies \ref{A:fluct}, as noted by \citet{breidt2017model}, since the conditional bound in \ref{A:fluct} reduces to the unconditional dependence control in \ref{A:depend}. For general $K > 1$, it is automatically satisfied for independent sampling under fold balance, and in Section \ref{sec:Controlling-the-remainder} we
verify that it also holds for several commonly used dependent sampling
designs.}

We now establish the following lemma. 
\begin{lemma}[Mean-square control of the Sample-split remainder]\label{lem:4.2}
Assume \ref{A:pos}, \ref{A:fluct}, and \ref{A:CF}. Fix $0<\eta<1$.
Then there exist constants $C_\eta<\infty$ and $N_\eta$ such that, with probability at least
$1-\eta$ for all $N\ge N_\eta$, for each $k=1,\dots,K$,
\begin{equation}
\Ep\Bigg[\Big(\frac{R_{N,k}}{N_{k}}\Big)^{2}\ \Big|\ \mathcal{G}_{k}\Bigg]
\ \le\ \frac{C_{\eta}}{n}\, \big\|\widehat{m}^{(-k)}-m\big\|_{N,k}^{2}.
\label{eq:foldwise-ms-bound}
\end{equation}
\end{lemma}
Lemma \ref{lem:4.2} generalizes Lemma \ref{lem:4.1} to dependent sampling designs satisfying \ref{A:fluct}.
An immediate consequence of Lemma \ref{lem:4.2} is that \ref{A:normcons} can
be used to show that the remainder term is negligible.

\begin{corollary}[Asymptotic equivalence to the oracle difference
estimator]\label{cor:ae_to_oracle}

Assume \ref{A:pos}, \ref{A:fluct}, \ref{A:CF}, and \ref{A:normcons}. Then 
\begin{equation}
\sqrt{n}\,\frac{R_{N}}{N}=o_{p}(1).\label{eq:15}
\end{equation}
Consequently, 
\begin{equation}
\sqrt{n}\,\frac{\widehat{T}_{\mathrm{SREG}}-T}{N}=\sqrt{n}\,\frac{\widehat{T}_{\mathrm{diff}}-T}{N}+o_{p}(1).\label{eq:16}
\end{equation}
\end{corollary}

We impose a standard design-based central limit theorem (CLT) for
the HT estimator applied to the residuals $e_{i}=y_{i}-m_{i}$.

\begin{enumerate}[label=(A\arabic*), leftmargin=*, start = 6] 
\item \label{A:CLT} \textbf{HT--CLT for the residual total.} Assume that 
\begin{equation}
\sqrt{n}\,\frac{\widehat{T}_{\mathrm{diff}}-T}{N}\ \xrightarrow{d}\ N(0,\,V),\label{eq:ht_clt_resid}
\end{equation}
where $V>0$ is the design-based asymptotic variance of the HT estimator
for $N^{-1}\sum_{i\in U_{N}}e_{i}$.
\end{enumerate}
Assumption \ref{A:CLT} holds in many classical samples under certain regularity conditions as described in Chapter 1 of \citet{fuller09}.
Combining (\ref{eq:16}) with (\ref{eq:ht_clt_resid})
and applying Slutsky's theorem, we obtain the following theorem.

\begin{theorem}[Asymptotic normality of the sample-split regression estimator]\label{thm:ss_clt}
Let $\widehat{T}_{\mathrm{SREG}}$ be the $K$-fold sample-split regression
estimator and $\widehat{T}_{\mathrm{diff}}$ be the oracle difference
estimator. Define the remainder $R_{N}:=\widehat{T}_{\mathrm{SREG}}-\widehat{T}_{\mathrm{diff}}$.

Assume \ref{A:pos}, \ref{A:fluct}, \ref{A:CF}, and \ref{A:normcons} so that Corollary~\ref{cor:ae_to_oracle}
holds. Further assume \ref{A:CLT}, a design-based CLT for the oracle difference
estimator. Then: 
\begin{equation}
\sqrt{n}\,\frac{\widehat{T}_{\mathrm{SREG}}-T}{N}\ \xrightarrow{d}\ N(0,V).\label{eq:thm1_clt_ss}
\end{equation}
\end{theorem}

We now discuss variance estimation. Let $\pi_{ij}$ be the joint inclusion
probability of units $i$ and $j$ in the population. For each fold
$k$, define out-of-fold residuals for sampled units $i\in A_{k}$:
\[
\widehat{e}_{i}^{(-k)}:=y_{i}-\widehat{m}_{i}^{(-k)}.
\]
A natural variance estimator is the HT variance estimator applied
to the cross-fitted residuals: 
\begin{equation}
\widehat{V}(\widehat{T}_{\mathrm{SREG}})=\sum_{k=1}^{K}\sum_{\ell=1}^{K}\sum_{i\in A_{k}}\sum_{j\in A_{\ell}}\frac{\pi_{ij}-\pi_{i}\pi_{j}}{\pi_{ij}}\cdot\frac{\widehat{e}_{i}^{(-k)}}{\pi_{i}}\cdot\frac{\widehat{e}_{j}^{(-\ell)}}{\pi_{j}}.\label{eq:ss_var_est}
\end{equation}

\begin{theorem}[Consistency of the variance estimator]\label{thm:var_consistency}

Assume, in addition
to \ref{A:pos}, \ref{A:depend}, \ref{A:fluct}, \ref{A:moments}, \ref{A:CF}, and \ref{A:normcons},
\begin{enumerate}[label=(A$'$\arabic*), leftmargin=*]
\item \label{A:pos2} (Positivity for second-order inclusion probabilities.) There exists $\pi_{2,\min}>0$
such that $\pi_{ij}\ge\pi_{2,\min}$ for all $i\neq j$ and all $N$. 
\setcounter{enumi}{2}
\item \label{A:moments2}(Fourth-moment bound for oracle residuals.) With $e_{i}=y_{i}-m_{i}$,
\[
\frac{1}{N}\sum_{i\in U_{N}}e_{i}^{4}=O(1).
\]
\setcounter{enumi}{4}
\item \label{A:normcons2} (Second-moment of Cross-fitted prediction error.) Let $\delta_{i}:=\widehat{m}_{i}^{(-k(i))}-m_{i}$.
Assume 
\[
\frac{1}{N}\sum_{i\in U_{N}}\delta_{i}^{2}=o_p(1).
\]
\setcounter{enumi}{6}
\item \label{A:varest} (Design consistency of the HT variance estimator.) Conditional
on any fixed finite-population values $\{a_{i}\}_{i\in U_{N}}$ with
$N^{-1}\sum_{i\in U_{N}}a_{i}^{4}=O(1)$, the HT
variance estimator 
\[
\widehat{V}_{\mathrm{HT}}(a):=\sum_{i\in A}\sum_{j\in A}\frac{\Delta_{ij}}{\pi_{ij}}\frac{a_{i}}{\pi_{i}}\frac{a_{j}}{\pi_{j}}
\]
satisfies 
\[
\frac{n}{N^{2}}\left\{ \widehat{V}_{\mathrm{HT}}(a)-\Varp\!\left(\sum_{i\in A}\frac{a_{i}}{\pi_{i}}\right)\right\} \xrightarrow{p}0.
\]
\end{enumerate}
Then the proposed variance estimator in \eqref{eq:ss_var_est} satisfies
\begin{equation}
\frac{n}{N^{2}}\left\{ \widehat{V}\!\left(\widehat{T}_{\mathrm{SREG}}\right)-\Varp\!\left(\widehat{T}_{\mathrm{diff}}\right)\right\} \xrightarrow{p}0,\label{eq:Vhat_consistency_oracle_scaled}
\end{equation}
where $\widehat{T}_{\mathrm{diff}}$ is the oracle difference estimator
using $m_{i}=m(\bm x_{i};\beta_{0})$. It follows that

\begin{equation}
\frac{\widehat{T}_{\mathrm{SREG}}-T}{\sqrt{\hat V(\hat T_{\rm SREG})}}\ \xrightarrow{d}\ N(0,1).
\end{equation}
\end{theorem}

\begin{remark}[Discussion of the theory]

Several features of the theory are worth emphasizing. First, the leading
remainder term depends only on the prediction accuracy of the out-of-fold
estimator, not on the estimation error of regression coefficients.
Second, the theory accommodates a wide class of regression and machine-learning
methods, provided that cross-fitted prediction error vanishes in mean
square over the finite population. Finally, the design-based CLT required
for inference is the same as that for the oracle difference estimator,
so no additional complexity is introduced by sample splitting.

\end{remark}

\section{Verification of the Conditional Fluctuation Condition}\label{sec:Controlling-the-remainder}

In Section \ref{sec:theory}, we have seen that \ref{A:fluct} is the key
assumption to make $R_{N}$ asymptotically negligible. In this section,
we demonstrate that \ref{A:fluct} is satisfied in many sampling designs.

\subsection{Simple random sampling without replacement (SRSWOR)}

Assume that the sampling design is SRSWOR of fixed size $n$ from
the finite population $U_{N}$ of size $N$. Then $\pi_{i}=\pi=n/N$
for all $i\in U_{N}$. Let $n_{k}:=\sum_{i\in U_{k}}I_{i}=|A\cap U_{k}|$
denote the realized within-fold sample size. 
Conditioning on $\mathcal{G}_k$ (so that $n_k=n-\sum_{j\notin U_k}I_j$ is fixed), the remaining sample to be drawn from $U_k$ is SRSWOR of size $n_k$.
The following proposition shows that SRSWOR design satisfies the conditional fluctuation bound assumption \ref{A:fluct}.

\begin{proposition} \label{prop:srs} Fix a fold $k\in\{1,\dots,K\}$.
Let $\{a_{i}^{(k)}:i\in U_{k}\}$ be any $\mathcal{G}_{k}$-measurable
array (possibly random through its dependence on $\mathcal{G}_{k}$).
Define 
\[
D_{k}:=\frac{1}{N_{k}}\sum_{i\in U_{k}}\Big(\frac{I_{i}}{\pi}-1\Big)a_{i}^{(k)}.
\]
Then, conditional on $\mathcal{G}_{k}$, 
\begin{equation}
\Ep\!\left(D_{k}^{2}\mid\mathcal{G}_{k}\right)\;\le\;\frac{1}{\pi^{2}N_{k}^{2}}\left\{ \frac{n_{k}N_{k}}{N_{k}-1}+(n_{k}-\pi N_{k})^{2}\right\} \cdot\frac{1}{N_{k}}\sum_{i\in U_{k}}\left(a_{i}^{(k)}\right)^{2},\label{SRS-1}
\end{equation}
whenever $N_{k}\ge2$.

Moreover, under the asymptotic regime $N\to\infty$, $n\to\infty$,
$n/N\to f\in(0,1)$, and with fixed $K$ and balanced folds (e.g.,
$\min_{k}N_{k}\asymp N$), the random multiplier in \eqref{SRS-1}
is $O_{p}(n^{-1})$. In particular, for any $\eta>0$ there exists
a constant $C_{\eta}<\infty$ such that for all sufficiently large
$N$, 
\[
\Pp\!\left(\Ep(D_{k}^{2}\mid\mathcal{G}_{k})\le\frac{C_{\eta}}{n}\cdot\frac{1}{N_{k}}\sum_{i\in U_{k}}(a_{i}^{(k)})^{2}\right)\ge1-\eta.
\]
Since \(K\) is fixed, the bound holds simultaneously for all \(k=1,\dots,K\) by a union bound, and therefore verifies Assumption \ref{A:fluct}.
\end{proposition}
Related bounds of the same order appear as a consequence of the Bernstein--Serfling inequalities of \citet{bardenet2015concentration} for SRSWOR, whereas our argument is conditional on $\mathcal G_k$.

\subsection{Stratified simple random sampling without replacement (Stratified
SRSWOR)}

Assume that the finite population is partitioned into $H$ strata:
\[
U_{N}=\bigcup_{h=1}^{H}U_{h},\qquad U_{h}\cap U_{h'}=\emptyset\ (h\neq h'),\qquad|U_{h}|=N_{h},\quad\sum_{h=1}^{H}N_{h}=N.
\]
A stratified SRSWOR design draws, independently across strata, a simple
random sample of fixed size $n_{h}$ from stratum $U_{h}$. The stratum
inclusion probability is $\pi_{i}=\pi_{h}:=n_{h}/N_{h}$ for $i\in U_{h}$, 
and the overall sample size is $n:=\sum_{h=1}^{H}n_{h}$.
For each stratum $h$, define the stratum--fold intersection and
its size $
U_{hk}:=U_{h}\cap U_{k}$, $N_{hk}:=|U_{hk}|$,
and the corresponding sampled set and its size 
$A_{hk}:=A\cap U_{hk}$, $n_{hk}:=|A_{hk}|=\sum_{i\in U_{hk}}I_{i}$. 
Conditioning on $\mathcal{G}_k$, the induced design on each $U_{hk}$ is SRSWOR of size $n_{hk}$.

\begin{proposition}[Stratified SRSWOR implies the fold-wise conditional
fluctuation bound] \label{prop:strat_srs} Fix a fold $k\in\{1,\dots,K\}$.
Let $\{a_{i}^{(k)}:i\in U_{k}\}$ be any $\mathcal{G}_{k}$-measurable
array (possibly random through its dependence on $\mathcal{G}_{k}$).
Define 
\[
D_{k}:=\frac{1}{N_{k}}\sum_{i\in U_{k}}\Big(\frac{I_{i}}{\pi_{i}}-1\Big)a_{i}^{(k)}=\frac{1}{N_{k}}\sum_{h=1}^{H}\sum_{i\in U_{hk}}\Big(\frac{I_{i}}{\pi_{h}}-1\Big)a_{i}^{(k)}.
\]
Assume $N_{hk}\ge2$ for all $h$ (which holds for all large $N$
under balanced folds). Then, conditional on $\mathcal{G}_{k}$, 
\begin{equation}
\Ep\!\left(D_{k}^{2}\mid\mathcal{G}_{k}\right)\;\le\;\frac{1}{N_{k}}\Bigg[\underbrace{\sum_{h=1}^{H}\frac{n_{hk}}{\pi_{h}^{2}(N_{hk}-1)}}_{=:V_{k}}\;+\;\underbrace{\sum_{h=1}^{H}\frac{(n_{hk}-\pi_{h}N_{hk})^{2}}{\pi_{h}^{2}N_{hk}}}_{=:B_{k}}\Bigg]\cdot\frac{1}{N_{k}}\sum_{i\in U_{k}}\left(a_{i}^{(k)}\right)^{2}.\label{STR-1}
\end{equation}

Moreover, under the asymptotic regime where $H$ and $K$ are fixed
and 
\[
\frac{N_{h}}{N}\to w_{h}\in(0,1),\qquad\frac{n_{h}}{N_{h}}\to f_{h}\in(0,1)\quad(h=1,\dots,H),
\]
and under a balanced fold partition within strata (e.g., $\min_{h,k}N_{hk}\asymp N_{h}$),
the random multiplier in \eqref{STR-1} is $O_{p}(n^{-1})$. In particular,
for any $\eta>0$ there exists a constant $C_{\eta}<\infty$ such
that for all sufficiently large $N$, 
\[
\Pp\!\left(\Ep(D_{k}^{2}\mid\mathcal{G}_{k})\le\frac{C_{\eta}}{n}\cdot\frac{1}{N_{k}}\sum_{i\in U_{k}}\left(a_{i}^{(k)}\right)^{2}\right)\ge1-\eta.
\]
Hence stratified SRSWOR satisfies Assumption \ref{A:fluct} (in the ``with
high probability'' sense). \end{proposition}

\subsection{Rejective (conditional Poisson) sampling}

We consider rejective sampling (also called conditional Poisson sampling)
of a fixed sample size $n$ from $U_{N}$. One convenient characterization
is the following: there exist odds weights $\{\omega_{i}>0\}_{i\in U_{N}}$
such that the design probability of selecting a sample $s\subset U_{N}$
with $|s|=n$ is 
\begin{equation}
\Pp(A=s)=\frac{1}{Z_{n}}\prod_{i\in s}\omega_{i},\qquad Z_{n}:=\sum_{\substack{t\subset U_{N}\\
|t|=n
}
}\prod_{j\in t}\omega_{j}.\label{R-0}
\end{equation}
Equivalently, if $\{X_{i}\}$ are independent Bernoulli variables
with odds $P(X_{i}=1)/P(X_{i}=0)=\omega_{i}$, then $A=\{i:X_{i}=1\}\mid\sum_{i}X_{i}=n$
has distribution \eqref{R-0}.

Let $\pi_{i}:=\Pp(I_{i}=1)$ denote the (rejective) first-order
inclusion probabilities and assume the boundedness condition \ref{A:pos}:
$0<\pi_{\min}\le\pi_{i}\le\pi_{\max}<1$ for all $i$ and all $N$.
Define the dispersion 
$d:=\sum_{i\in U_{N}}\pi_{i}(1-\pi_{i})$. 
Under \ref{A:pos} and $n/N\to f\in(0,1)$, we have $d\asymp N\asymp n$.

Let $\{a_{i}^{(k)}:i\in U_{k}\}$ be any $\mathcal{G}_{k}$-measurable
array and define 
$D_{k}:=N_{k}^{-1}\sum_{i\in U_{k}}\Big(\pi_i^{-1} I_{i}-1\Big)a_{i}^{(k)}$. 
Conditional on \(\mathcal{G}_k\), the induced design on \(U_k\) remains rejective with fixed size \(n_k\), so standard second-moment bounds for rejective sampling apply fold-wise. 

\begin{proposition}[Rejective sampling implies the fold-wise conditional
fluctuation bound] \label{prop:rejective} Fix a fold $k\in\{1,\dots,K\}$.
Assume rejective sampling \eqref{R-0}, \ref{A:pos}, $n/N\to f\in(0,1)$,
and balanced folds (e.g., $cN\le N_{k}\le CN$ for constants $0<c<C<\infty$
and all large $N$). Then there exists a constant $C_{\eta}<\infty$
such that, for any $\eta>0$ and all sufficiently large $N$, 
\[
\Pp\!\left(\Ep(D_{k}^{2}\mid\mathcal{G}_{k})\le\frac{C_{\eta}}{n}\cdot\frac{1}{N_{k}}\sum_{i\in U_{k}}\big(a_{i}^{(k)}\big)^{2}\right)\ge1-\eta.
\]
In particular, rejective sampling satisfies Assumption \ref{A:fluct} (in
the ``with high probability'' sense). \end{proposition}
A related tail bound without $K$-splits was also established by \citet{bertail2019bernstein} under rejective sampling.

\section{Empirical Study}\label{sec:Simulation-Study}

\subsection{Simulation}
We conduct a simulation study to evaluate the finite-sample performance
of the proposed sample-split regression estimator (SREG) relative
to standard model-assisted estimators in a high-dimensional setting.
The focus is on three aspects emphasized by the theory: (i) bias under
informative sampling, (ii) efficiency trade-offs induced by sample
splitting, and (iii) the accuracy of variance estimation and confidence-interval
coverage.


A population $U$ of size $N=1000$ is generated from the following
scheme: $\bm{x}_{i},\ i=1,\cdots,N$ is a $p(=90)$-dimensional auxiliary
variable generated independently from  $MVN(\bm{\mu},\bm{\Sigma})$
where $\mu=(2,\cdots,2)^{T}$ and $(i,j)$-th component of $\Sigma$
is $0.2^{\abs{i-j}}$. The response variable $y_{i}$
is generated from the following super-population model: 
\[
y_{i}=\bm{x}_{i}^{T}\bm{\beta}+e_{i},\quad e_{i}\stackrel{iid}{\sim}N(0,\sigma^{2}),i=1,\cdots,N
\]
where $\sigma^{2}=1$ and ${\bm \beta=(\underbrace{1,\cdots,1}_{s=5},\underbrace{0,\cdots,0}_{p-s})^{T}}$.
The parameter of interest is the population total $T=\sum_{i\in U}y_{i}$. 


Samples of size $n=300$ are repeatedly drawn using a stratified simple
random sampling design. Informativeness is introduced through an auxiliary
variable $z_i = r e_i + \sqrt{1 - r^2}z_i^* $, where $z_i^* \sim N(0, 1)$ independent from $e_i$. $r(=-0.75)$ controls the strength of informativeness.
The population is sorted by $z_{i}$ and partitioned into four strata of equal size
($N_{h}=N / 4$), with stratum-specific sampling fractions
$(n_{1},n_{2},n_{3},n_{4})=(0.15n,0.20n,0.30n,0.35n)$. 
We also considered rejective sampling with fixed sample size $n=300$.
In this case, first--order inclusion probabilities $\pi_i$ are constructed
to be proportional to a positive size measure derived from the auxiliary
variable $z_i$, specifically $\pi_i \propto 1 / (1 + \exp(-z_i))$, and scaled so
that $\sum_{i=1}^N \pi_i = n$ with $0 < \pi_i < 1$.
A rejective (conditional Poisson) sampling scheme is then applied, whereby
independent Bernoulli samples with probabilities $\pi_i$ are repeatedly
drawn and retained only when the resulting sample size equals $n$.
This produces an unequal--probability sampling design with fixed sample
size, preserving informativeness through the dependence of $\pi_i$ on $z_i$.
Each simulation is replicated $B=500$ times. From each sample, we computed the following estimators. 
\begin{enumerate}
\item \textbf{HT}: Horvitz--Thompson estimator.
\item \textbf{Diff}: Oracle difference estimator using the true mean function \eqref{eq:oracle_diff}.
\item \textbf{GREG.Oracle}: GREG estimator \eqref{eq:greg} using only covariates with nonzero coefficients.
\item \textbf{GREG}: GREG estimator \eqref{eq:greg} with least-squares regression fit for $\bm{\beta}$ using all $p$ covariates.
\item \textbf{SREG}: Proposed SREG estimator \eqref{eq:TSS} with least-squares regression fit and $K=10$ folds.
\item \textbf{GREG.Lasso}: GREG estimator \eqref{eq:greg} with lasso regression fit for $\bm{\beta}$.
\item \textbf{SREG.Lasso}: Proposed SREG estimator \eqref{eq:TSS} with lasso regression fit and $K=10$ folds.
\end{enumerate}






Table \ref{tabref} presents the bias, standard error (SE), and root-mean-squared error (RMSE) of the point estimators. Under informative sampling, the GREG estimator exhibits substantial bias, despite correct model specification. In contrast, the proposed sample-split estimators (SREG and SREG.Lasso) effectively eliminate this bias and closely track the oracle difference estimator. Sample splitting introduces a modest increase in variance relative to ordinary GREG, as expected. However, when combined with penalization (SREG.Lasso), the variance inflation is substantially mitigated, yielding estimators with both low bias and competitive RMSE.

\begin{figure}[ht]
    \centering
    \includegraphics[width=\linewidth, height=5cm]{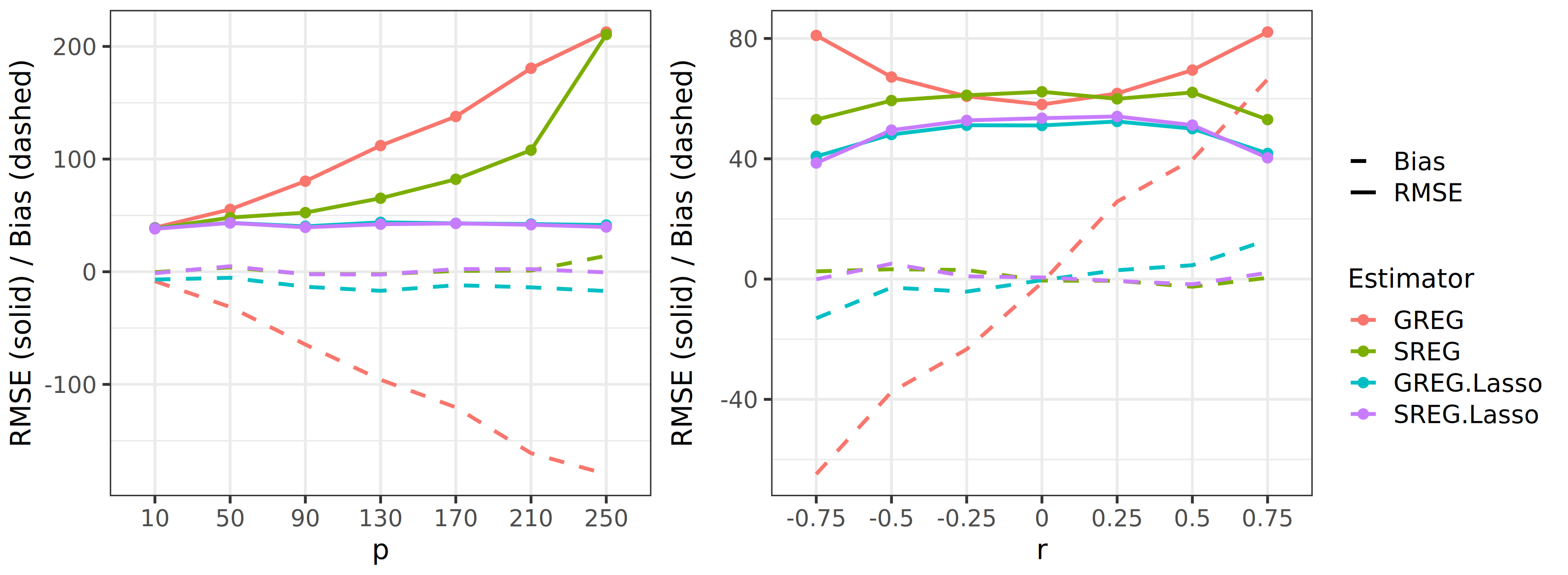}
    \caption{RMSE(solid line) and Bias(dashed line) for different values of $p$ (left) and $r$ (right) under stratified sampling. As \(p\) increases or \(|r|\) moves away from 0, the GREG estimator’s bias grows sharply, whereas the SREG estimator’s bias remains stable.}
    \label{fig1}
\end{figure}

Figure \ref{fig1} presents the bias and RMSE of the estimators across different values of $p$ and $r$ under stratified sampling.
GREG estimators are approximately unbiased, including
ordinary GREG under non-informative sampling ($r = 0$) or under low-dimensional variables ($p = 10$). It also shows that the bias of the ordinary GREG estimator increases sharply with sampling informativeness and high-dimensionality, whereas the bias of SREG remains stable. 

Table \ref{tabref2} reports results for relative bias (RB) of the variance estimator and the coverage rate (CR) of the 95\% confidence-interval.  Under informative sampling, variance estimators
paired with biased GREG estimators exhibit noticeable under-coverage.
In contrast, the proposed variance estimator applied to SREG shows
negligible relative bias (typically below $5\%$) and coverage rates
close to the nominal level. These findings are consistent with the
theoretical results in Section \ref{sec:theory}, which establish
consistency of the variance estimator once the leading bias is removed.
\subsection{Real-data application}
A real-data application provided in the supplementary material demonstrates  the practical utility and advantages of the proposed framework.

\section{Concluding Remarks}

\label{sec:Discussion}

This paper addresses a fundamental challenge in modern survey sampling:
how to conduct model-assisted regression estimation when auxiliary
information is high-dimensional.
We show that, in this regime, the ordinary GREG estimator can lose
its oracle-equivalence property and exhibit first-order bias, even
when the working regression model is correctly specified. The source
of this failure is structural due to the double use of sampled outcomes
in both regression fitting and residual correction, rather than a
consequence of variance inflation or model misspecification alone.

To resolve this issue, we propose a sample-split regression
(SREG) estimator that enforces an honesty constraint between prediction
and residual correction. By pairing each residual with an out-of-fold
prediction, the proposed estimator eliminates the leading self-influence
term responsible for high-dimensional bias. The resulting discrepancy
relative to the oracle difference estimator becomes a HT fluctuation
applied to the out-of-fold prediction error, which can be controlled
under a weak prediction-norm consistency requirement.

A key strength of the proposed approach is its flexibility. The theoretical
results do not require $\sqrt{n}$-consistent estimation of regression
coefficients, nor do they impose sparsity or rely on the correctness of the working model. Any regression or prediction method that delivers
out-of-fold consistency in prediction norm over the finite population
is admissible.  This feature positions the SREG estimator
as a natural bridge between traditional design-based survey inference
and modern predictive modeling. The framework uses prediction accuracy as the primary
requirement and leverages design-based theory to guarantee valid inference.

This work opens several directions for future research. First, while
we verify the key conditional fluctuation condition for several widely
used sampling designs, extending the analysis to more complex  designs, such as general rejective sampling \citep{yang2025two},
would be valuable. Second, how to implement the SREG estimator using generalized entropy weight calibration \citep{kwon2025debiased} will be an interesting research topic. 
Finally, while our analysis is design-based, it would be interesting to explore hybrid perspectives that combine design-based guarantees with superpopulation modeling, particularly in settings where auxiliary information is learned from external data sources.


\section*{Data Availability}
All datasets and code used in this paper can be found at \url{https://github.com/yonghyun-K/SREG}.





\bibliographystyle{abbrvnat}
\bibliography{ref}

\begin{table}[htbp]
\centering %
\caption{Bias, SE, and RMSE under stratified (left) and rejective (right) sampling. Ordinary GREG is biased under informative sampling; sample splitting removes the bias (SREG), and penalization curbs variance inflation (SREG.Lasso), giving RMSE near the oracle (GREG.Oracle).}
\begin{tabular}{rrrr}
\multicolumn{4}{l}{Stratified sampling}\tabularnewline
\hline 
 & BIAS & SE & RMSE\tabularnewline
\hline 
HT & 6.55 & 148.05 & 148.20 \\ 
  Diff & -2.04 & 36.68 & 36.74 \\ 
  GREG.Oracle & -5.69 & 36.98 & 37.42 \\ 
  GREG & -67.18 & 48.48 & 82.85 \\ 
  SREG & \bf -0.90 & 51.68 & 51.69 \\ 
  GREG.Lasso & -13.35 & 38.75 & 40.99 \\ 
  SREG.Lasso & \bf -1.07 & 39.86 & 39.87 \\ 
\hline 
\end{tabular}
\begin{tabular}{rrrr}
\multicolumn{4}{l}{Rejective sampling}\tabularnewline
\hline 
 & BIAS & SE & RMSE\tabularnewline
\hline 
HT & 20.63 & 397.60 & 398.14 \\ 
  Diff & 0.41 & 68.59 & 68.59 \\ 
  GREG.Oracle & -5.49 & 76.82 & 77.01 \\ 
  GREG & -93.95 & 75.53 & 120.55 \\ 
  SREG & \bf 0.56 & 88.86 & 88.87 \\ 
  GREG.Lasso & -16.33 & 79.90 & 81.55 \\ 
  SREG.Lasso & \bf 3.18 & 82.74 & 82.80 \\ 
\hline 
\end{tabular}
\label{tabref}
\end{table}


\begin{table}[htbp]
\centering %
\caption{Relative bias (RB) and 95\% CI coverage (CR) under stratified (left) and rejective (right) sampling. Under informative sampling, GREG-type methods yield under-coverage, whereas SREG (and SREG.Lasso) has small RB and near-nominal coverage.}

\begin{tabular}{rrr}
\multicolumn{3}{l}{Stratified sampling}\tabularnewline
\hline 
 & RB  & CR \tabularnewline
\hline 
  GREG & -0.49 & 0.53 \\ 
  SREG &  0.01 & \bf 0.94 \\ 
  GREG.Lasso & -0.12 & 0.91 \\ 
  SREG.Lasso & -0.02 & \bf 0.94 \\ 
\hline 
\end{tabular}
\begin{tabular}{rrr}
\multicolumn{3}{l}{Rejective sampling}\tabularnewline
\hline 
 & RB  & CR \tabularnewline
\hline 
  GREG & -0.35 & 0.59 \\ 
  SREG & -0.02 & \bf 0.93 \\ 
  GREG.Lasso & 0.00 & 0.90 \\ 
  SREG.Lasso & 0.09 & \bf 0.93 \\ 
\hline 
\end{tabular}
\label{tabref2} 
\end{table}

\end{document}